\documentclass[aps,prb,twocolumn,superscriptaddress,floatfix,showpacs]{revtex4}
\usepackage{graphicx}
\usepackage{dcolumn}
\usepackage{latexsym}
\usepackage{amssymb}
\usepackage{amsfonts}
\usepackage{amsmath}
\usepackage{fancybox}
\usepackage[binary,squaren]{SIunits}
\usepackage{colordvi}
\usepackage[normalem]{ulem}
\usepackage{cancel}

\def\be{\begin{equation}}
\def\ee{\end{equation}}
\def\ber{\begin{eqnarray}}
\def\eer{\end{eqnarray}}

\def\rv{{\bf r}}

\def\pv{{\bf p}}

\def\kv{{\bf k}}

\def\ep{{\epsilon}}

\begin{document}
\title{Spin current swapping and Hanle spin Hall effect in the two dimensional electron gas}
\author{Ka Shen}
\affiliation{Department of Physics and Astronomy, University of Missouri,
  Columbia, Missouri 65211, USA}
\affiliation{Kavli Institute of NanoScience, Delft University of Technology, Lorentzweg
1, 2628 CJ Delft, The Netherlands}
 \author{R. Raimondi}
\affiliation{CNISM and Dipartimento di Matematica e Fisica, Universit\`a Roma Tre,
Via della Vasca Navale 84, 00146 Rome, Italy}

\author{G. Vignale}
\affiliation{Department of Physics and Astronomy, University of Missouri,
  Columbia, Missouri 65211, USA}
\pacs{72.25.-b, 71.70.Ej, 72.20.Dp, 85.75.-d}


\date{\today}
\begin{abstract}
We analyze the effect known as ``spin current swapping" (SCS) due to electron-impurity scattering in a uniform spin-polarized two-dimensional electron gas.  In this effect a primary spin current $J_i^a$ (lower index for spatial direction, upper index for spin direction) generates a secondary spin current $J_a^i$ if $i \neq a$, or $J_j^j$, with $j\ne i$,  if $i= a$.     Contrary to naive expectation,  the homogeneous spin current associated with the uniform drift of the spin polarization in the electron gas  does not generate a swapped spin current by the SCS mechanism.   Nevertheless, a swapped spin current will be generated, if a magnetic field is present,  by a completely different mechanism, namely, the precession of the spin Hall spin current in the magnetic field.    We refer to this second mechanism as {\it Hanle spin Hall effect}, and we notice that it can be observed in an experiment in which a homogeneous drift current is passed through a uniformly magnetized electron gas.   In contrast to this, we show  that  an unambiguous  observation of SCS requires inhomogeneous spin currents, such as those that are associated with spin diffusion in a metal, and no magnetic field.  An experimental setup for the observation of the SCS is therefore proposed.
\end{abstract}
\maketitle

\section{Introduction} 
\begin{figure}
\includegraphics[width=1.65in]{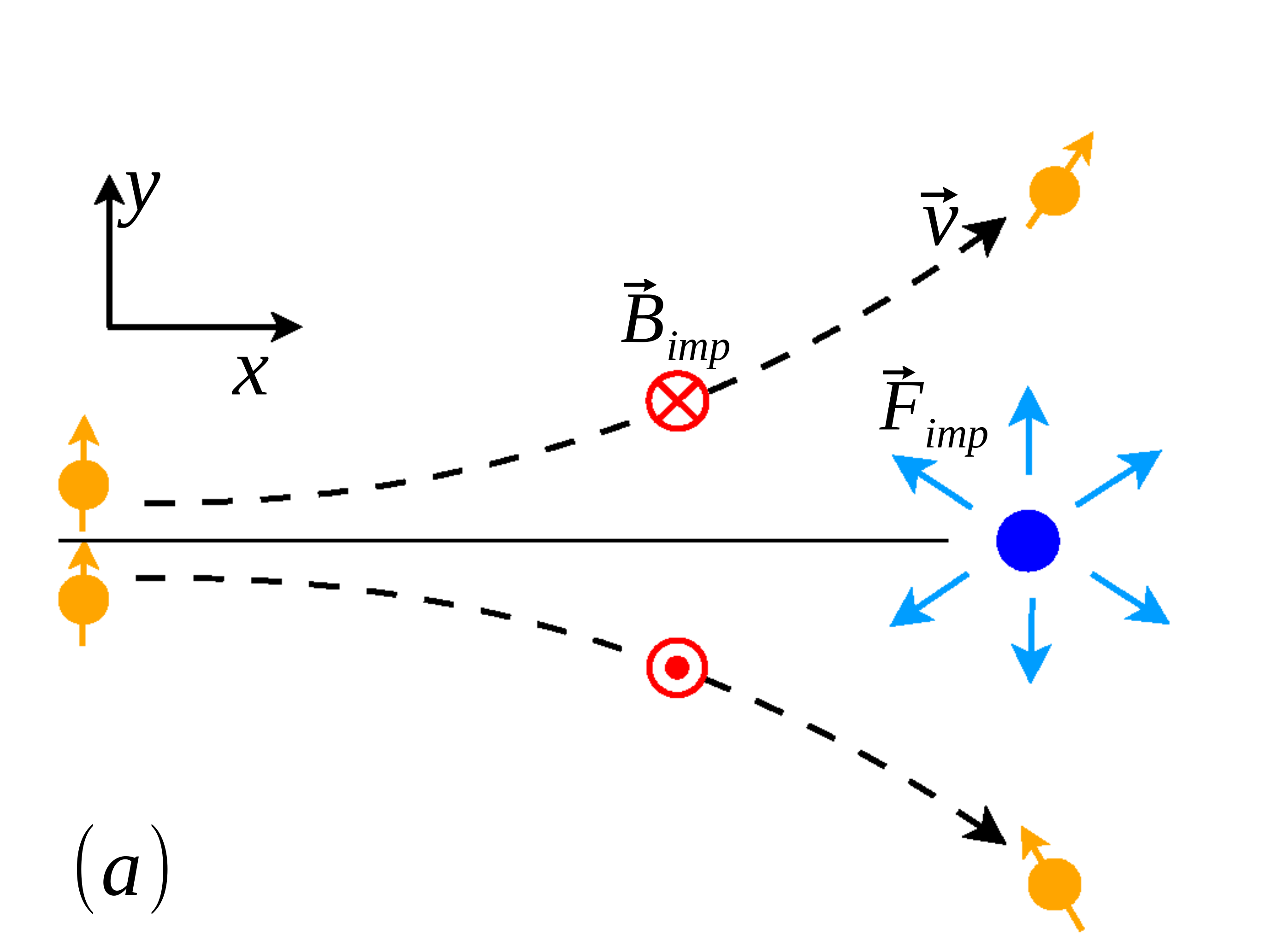}
\includegraphics[width=1.65in]{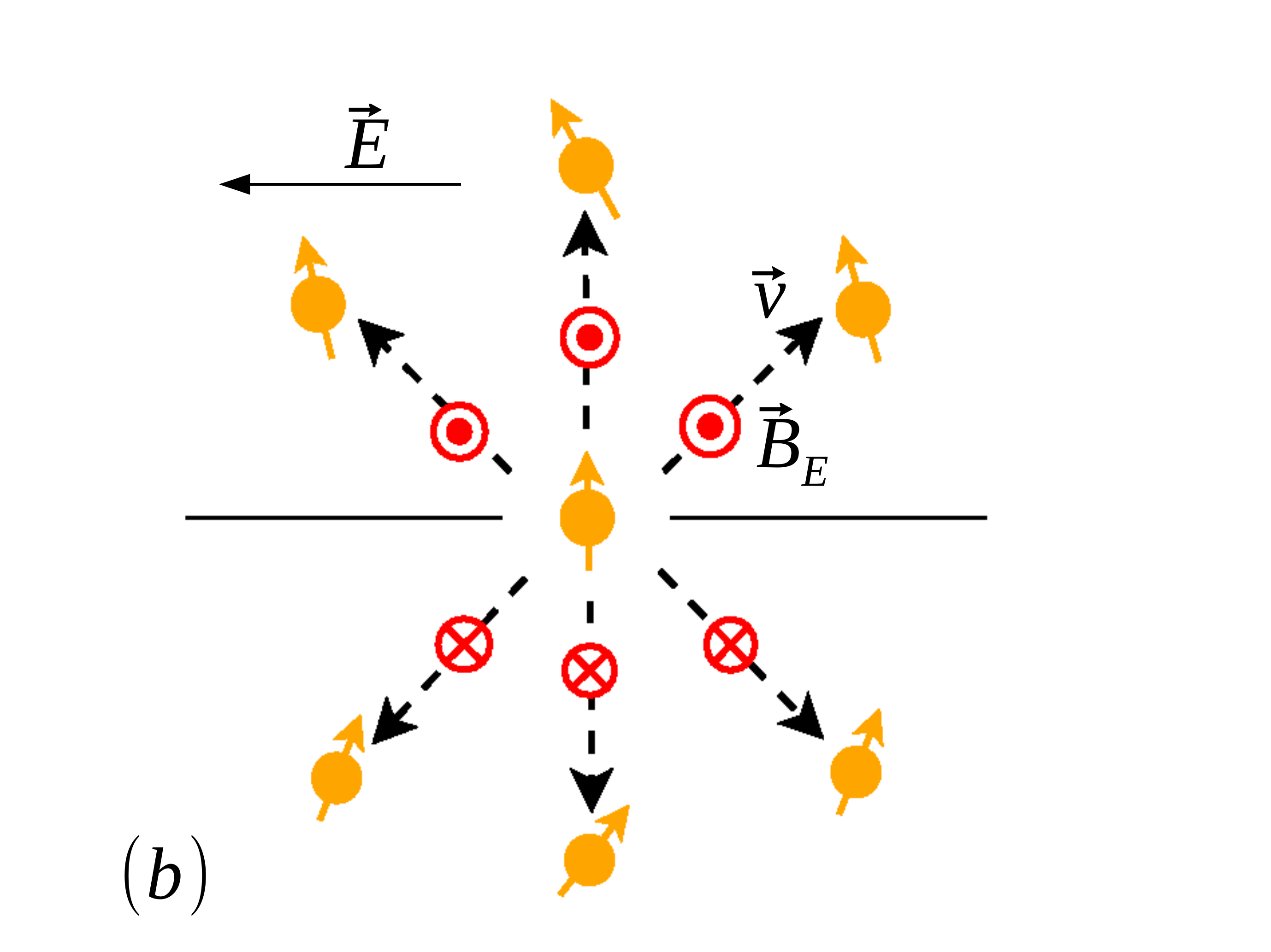}
\caption{(a) Impurity-induced spin current swapping for primary spin current $J_x^y$. The electric force induced by the impurity potential ($-\nabla V_{\rm imp}$ shown as the blue arrows) gives rise to an effective magnetic field ${\vec B}_{\rm imp}\simeq \lambda^2 m \vec F_{\rm imp} \times \vec v$, which, so as to the spin precession, is in opposite directions for the two trajectories, leading to a transverse spin current $J_y^x$. (b) The spin current swapping effect due to the external electric field. Here, the spin precession is caused by the effective magnetic field ${\vec B}_E\simeq -\lambda^2 me  \vec E\times \vec v$. Observe how the ``swapped" spin current $J^x_y$ produced by the electric field in (b)  is opposite to the one produced by the impurities in (a).
}
\label{phys_pic}
\end{figure}

The generation, manipulation and detection of spin currents are central issues in realizing spintronic devices.\cite{rmp_76_323,Awschalom07np} Recently, Lifshits and D'yakonov~\cite{Lifshits09} described an interesting and potentially important ``spin current swapping" (SCS) effect: a primary spin current, $[J_a^i]^{(0)}$ flowing along the $a$ direction with spin polarization along the $i$ direction, generates a transverse spin current, which can be expressed  as
\be
[J_i^a]^{\rm SCS}=\kappa\left([J_a^i]^{(0)}-\delta_{ia}\sum_l [J_l^l]^{(0)}\right),\label{SCS1}
\ee
with the generation efficiency parameter $\kappa =\lambda^2 k_F^2$ proportional to the square of the effective Compton wavelength $\lambda$ (which controls the strength of the spin-orbit coupling) and the square of the Fermi wave vector $k_F$.  Here the lower index, $i$, denotes the spatial direction of  flow of the spin current, while the upper index $a$ denotes the orientation of the spin.    As discussed by Lifshits and D'yakonov in Ref.~\onlinecite{Lifshits09}, the SCS effect originates from the spin precession of the propagating electrons under the impurity-generated spin-orbit field.  In a classical picture, when an electron passes near an impurity, not only its momentum changes, but also its spin  undergoes a rotation around the effective magnetic field associated with the impurity potential.  This effective field is normal to the plane defined by the electron momentum and the gradient of the local electric potential, and its sign depends on whether the electron passes on the left or on the right side of the impurity.  The correlation between the scattering direction and the sign of the spin precession is the essence of the spin-current swapping effect, as it causes, for example, spins initially oriented in the $+y$ direction and propagating along $+x$ (primary spin current $J_x^y$) to acquire a positive $x$ spin component when they are deflected in the positive $y$ direction,  and a negative $x$ component when they are deflected in the negative $y$ direction: this results in a secondary spin current $J_y^x$ [see Fig.~\ref{phys_pic}(a) and its caption].

At first sight the detection of the spin current swapping effect seems quite straightforward.  Consider, for example, a setup consisting of a two dimensional electron gas confined to the $x$-$y$ plane, with a spontaneous or induced in-plane spin polarization described by average homogeneous spin densities $S^x$ and $S^y$.    An electric field applied in the $-x$ direction will produce  primary spin currents  $[J_x^x]^{(0)}$ and $[J_x^y]^{(0)}$  proportional to the charge current $J_x$ and to the spin densities  $S^x$ and $S^y$, respectively.   The effective magnetic field created by the spin-orbit interaction  with the impurities  is perpendicular to the plane, i.e., along the $z$ direction.  Then, according to Eq.~(\ref{SCS1}), the spin currents generated by the spin current swapping effect should be
\ber\label{SCSeffect}
 [J_y^y]^{\rm SCS} &=& -\kappa [J_x^x]^{(0)},\\
{[J_y^x]}^{\rm SCS} &=& \kappa [J_x^y]^{(0)}\,
\eer
and it might seem a relatively easy matter to detect the spin accumulations associated with one or the other component of the spin current.
Notice that  both $[J_y^y]^{\rm SCS}$ and ${[J_y^x]}^{\rm SCS}$ are transverse  with respect to the direction of flow of the primary current.   However, at variance with the well-known transverse spin current induced by spin Hall effect $J_y^z$~(Refs.~\onlinecite{DPshe71,Sinova04,Kato04,Wunderlich05}), here only the in-plane spin components are relevant.

Unfortunately, things are not so simple. As explained above, Eq.~(\ref{SCSeffect})  takes into account only  the effect of the out-of-plane magnetic field from impurity spin-orbit coupling.   The  in-plane external electric field that drives the primary spin current  -- a plain drift current --  will also contribute to the SCS, because it generates, via spin-orbit coupling,  an effective magnetic field that lies exactly in the opposite direction as the impurity-induced one [see Fig.~\ref{phys_pic}(b)].   Therefore, the total SCS spin current will be the sum of two contributions, one from the impurities and the other from the electric field, and these two contributions cancel each other.   This point was recognized and discussed in our recent paper,\cite{Shenprb14} where we demonstrated the exact cancellation (in a homogeneous system) from spin-dependent drift-diffusion equations. The cancellation can be understood as a consequence of the force balance between external electric field and impurities at the steady state, i.e., $\langle \nabla V_{\rm imp}\rangle= \vec E$, with $\langle ...\rangle$ representing the average over the electron density distribution.

\begin{figure}
\includegraphics[width=2in]{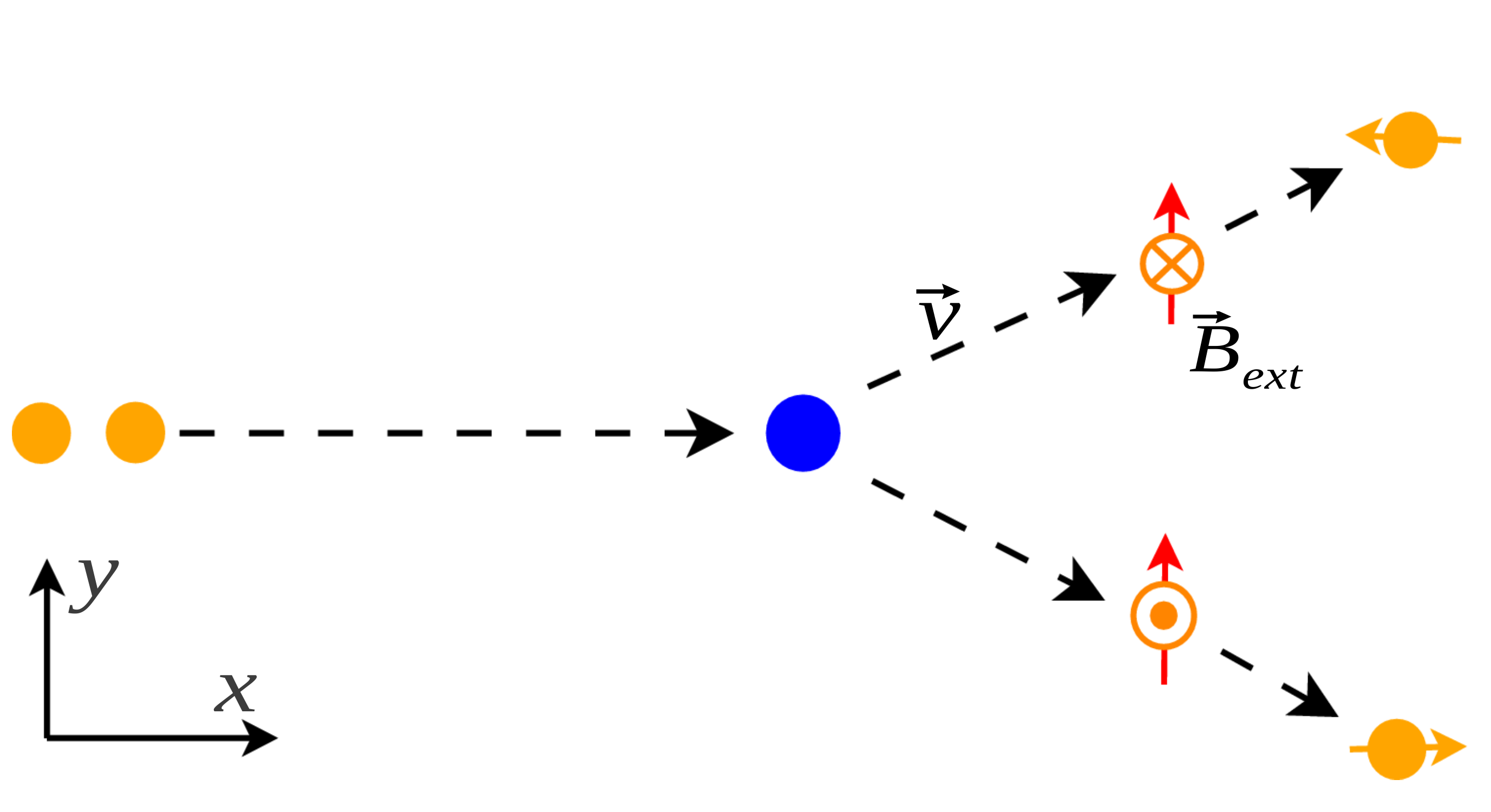}
\caption{Schematics of the Hanle spin Hall effect.  A charge current along the $x$-axis generates a spin Hall current $J^z_y$.  In the time $\tau$ between collisions the spins, initially pointing in the $z$ directions, precess in the magnetic/exchange field (red arrows), acquiring a finite $x$-component  if the field is along $y$ (case shown in the figure), or an $y$-component, if the field is along $x$ (not shown).  Collisions with impurities, ``reset" the orientation of the spin along the $z$ axis.}
\label{Hanle-SHE}
\end{figure}

The analysis of Ref. \onlinecite{Shenprb14}  was, however, incomplete, because it did not take into account the action of the external magnetic field and/or the internally generated spin-dependent potential (exchange potential), which may be responsible for the spin-polarization of the electron gas.  
While a spin polarization can exist, out of equilibrium, even without a magnetic field, it is important to understand how the generation of spin currents and the  SCS will be affected by the presence of such a field.   In this paper  we provide an answer to this question.  Specifically, we point out the existence of an effect that can easily pass for SCS even though its physical origin is quite different.    This effect, illustrated in Fig.~\ref{Hanle-SHE}, arises from the combined action of the spin Hall effect, which generates a spin current $J^z_y$ when the electric current is in the $x$ direction,  and the spin precession driven by the external magnetic field and/or the exchange-field of the ferromagnet, which rotates $J^z_y$   around the axis of the spin polarization producing a small ``swapped" spin current  $J^x_y$, if the magnetic/exchange  field is in the $y$ direction (red arrows), or $J^y_y$, if that field is in the $x$ direction.   Due to its similarity with the well known Hanle effect, in which a non equilibrium spin polarization is rotated by a magnetic field away from its original direction, we call this effect ``Hanle spin Hall effect" (HSHE).  What is rotated is the spin Hall current, resulting in the generation of a secondary spin current which is virtually indistinguishable from SCS. 

In Section II of this paper we present the diagrammatic theory of the HSHE in a disordered spin polarized two-dimensional electron gas.  Both side-jump and skew-scattering contributions to the spin Hall current are considered (the latter in Appendix B) and we show that both give rise to SCS-like spin currents when an in-plane magnetic field is present.  The gedanken experiment described at the beginning of this introduction would therefore be an experimental test not of the spin current swapping but of the  Hanle spin Hall effect  

Next, in Section III we show that the SCS effect can be observed in inhomogeneous situations, such as the one described in Ref.~\onlinecite{Lifshits09} where the spin current was injected from ferromagnetic leads into a nonmagnetic conductor.  In this case the primary spin current is a {\it diffusion current}, driven by a spin density gradient rather than by an electric field, and our arguments leading to the cancellation of the SCS for drift currents do not apply.  We therefore propose an experimental setup for the observation of SCS in a nonmagnetic metal.

Lastly, in Section IV we clarify the relation between the SCS discussed in this paper -- clearly an effect arising from spin-orbit interaction with  impurities -- and the ``intrinsic SCS" introduced by Sadjina {\it et al.} in Ref.~\onlinecite{Sadjina12}.

\section {Diagrammatic theory of HSHE} We consider a homogeneous two dimensional electron gas with a finite homogeneous spin polarization along the $x$ axis:  the polarization is maintained by an external magnetic field or by an internal exchange field with, say, $d$-electrons.   A longitudinal electric field $E_x$ produces a drift current of charge ($J_x$) and spin ($J_x^x$).  We use the standard Kubo formula to calculate the transverse spin swapping current ($J_y^y$) in the presence of spin-orbit coupling with  impurities. Our model Hamiltonian is
\be
\label{Hamiltonian}
H=\hat p^2/(2m)+V(\mathbf r)-({\Delta}/{2}) \hat \sigma^x -\lambda^2 \vec{\hat\sigma}\times\nabla V(\mathbf r)\cdot \hat{\mathbf p},
\ee
with $\hat {\bf p}=-i\nabla_{\bf r}$ and $V(\mathbf r)$ representing a short-range impurity potential with zero average and Gaussian distribution given by
$\langle V({\mathbf r})V({\mathbf r'})\rangle =v_0^2\delta ({\mathbf r} -{\mathbf r'})$. Notice that we have set $\hbar=1$.  Here $\Delta$ is the difference of the Fermi energies, $E_+$ and $E_-$ of the two spin bands with $\sigma^x=\pm 1$: $\Delta=E_+-E_-$. 
Within the self-consistent Born approximation,
the retarded and advanced Green's functions have the form
\begin{equation}
\label{11}
\hat G^{R/A}_{\bf k}(\epsilon) =\hat\sigma^0G^{R/A}_{0\bf k}(\epsilon)+\hat\sigma^x G^{R/A}_{1\bf k}(\epsilon),
\end{equation}
where 
\begin{eqnarray}
G^{R/A}_{0\bf k} & = & \frac{1}{2}( G_{+\bf k}^{R/A}+G_{-\bf k}^{R/A}), \\
G^{R/A}_{1\bf k} & = & \frac{1}{2}( G_{+\bf k}^{R/A}-G_{-\bf k}^{R/A}),
\end{eqnarray}
with $G^R_{\pm \mathbf k}(\epsilon)=(\ep -\xi_{\kv}\pm\Delta /2 +{\rm i}/2\tau)^{-1}$ and $G^A_{\pm \mathbf k}=(G^R_{\pm \mathbf k})^\ast$. Here, $\xi_{\mathbf k}=k^2/(2m)-E_F$ with the average Fermi energy $E_F=(E_++E_-)/2$, and  $\hat\sigma^i$ (with $i=0,x,y,z$) are the usual  Pauli matrices.  The scattering time has the standard expression $\tau^{-1}=2\pi n_i N_0 v_0^2$, with $N_0=m/2\pi$ and $n_i$ being the density of states and impurity concentration in two dimensions, respectively.
Notice that, in using the self-consistent Born approximation, we have absorbed the $\hat\sigma^0$ and $\hat\sigma^x$ components of the real part of the Green function self-energy into the renormalization of the chemical potential and Zeeman energy $\Delta$, respectively (See Ref. \onlinecite{schwab2002} for details).

According to the linear response theory, the longitudinal and transverse spin currents arising from the application of an electric field along the $x$ axis are
\begin{equation}
\label{Jxx}
J^x_x=\sigma^x_{xx} E_x,
\end{equation}
and 
\begin{equation}
\label{Jyy}
 J^y_y=\sigma^{y}_{yx} E_x,
\end{equation}
where $\sigma^x_{xx}$ and $\sigma^y_{yx}$ are the longitudinal and transverse spin conductivities, respectively.
\begin{figure}
\begin{center}
\includegraphics[width=2in]{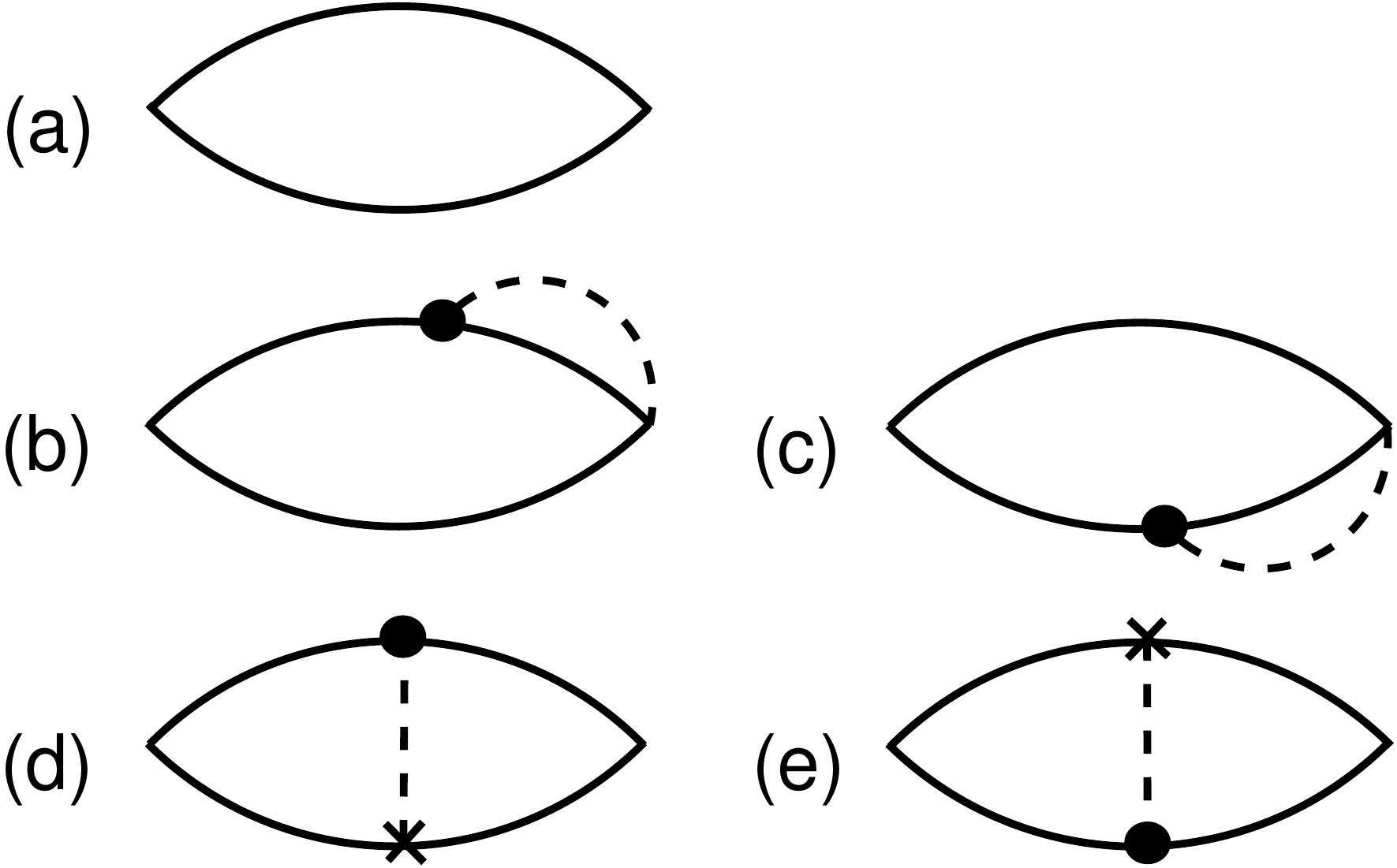}
\caption{Diagrams for the evaluation of the swapping coefficient when time-reversal symmetry is broken by a non-vanishing spin polarization. The left and right vertices are spin current vertex ($\hat J_x^x$ or $\hat J_y^y$) and charge current one ($\hat J_x$), respectively. {The solid lines are Green's functions including the standard impurity self-energy, and the dashed lines represent the impurity potential correlator.} (a) Diagram contributing to $\sigma^{x}_{xx}$.
 (b) and (c)  Side-jump type of diagrams contributing to $\sigma^{y}_{yx}$ originating from the anomalous velocity
operator $\delta \hat v_x$ defined in the text.  (d) and (e) vertex corrections contributing to $\sigma^{y}_{yx}$  to first order in the spin-orbit coupling. The cross denotes the spin-orbit from impurity potential, whereas the full dot is the standard (spin-independent)  impurity potential.
}
\label{diagram_2}
\end{center}
\end{figure}
The longitudinal spin conductivity is given by  
\be
\sigma^x_{xx}=\lim_{\omega\to 0}\frac{\langle\langle \hat J_x^x;\hat J_x\rangle\rangle_\omega}{i\omega}\,,
\ee
where the double bracket denotes the  Kubo product $\langle\langle \hat A;\hat B\rangle\rangle_\omega\equiv -\frac{i}{\hbar}\int_0^t \langle[\hat A(t),\hat B(0)]e^{i\omega t}dt$.  The zero-th  diagram, shown in Fig.~\ref{diagram_2}(a), gives
\begin{equation}
\label{14}
\sigma_{xx}^x=\frac{1}{2\pi}\sum_{\kv} {\rm Tr} \left( \hat J^x_x\hat G^R \hat J_x\hat G^A\right),
\end{equation}
where we have introduced charge-current and spin-current vertices as
\begin{eqnarray}
\hat J_x & = & (-e)\hat v_x \\
\hat J^x_x & = &  \frac{1}{2}\hat v_x \hat\sigma^x,
\end{eqnarray}
with velocity operator $\hat v_x={k_x}/{m}$. Note that $e$ is the positive unit charge and we assign to electrons a charge $-e$. By performing the integral over momentum, we get
\begin{equation}
\label{15}
\sigma_{xx}^x=(-e) \frac{N_0 D}{2}\frac{2 \Delta}{mv_F^2},
\end{equation}
where 
$D=v_F^2\tau /2$ is the diffusion coefficient with $v_F=\sqrt{2E_F/m}$ being the Fermi velocity. By noting that the difference between the squares of the Fermi momenta of the two Fermi surfaces is $k^2_{F+}-k^2_{F-}=2m \Delta$, Eq.~(\ref{15}) can also be written as
\begin{equation}
\label{16}
\sigma_{xx}^x=(-e)\frac{1}{2}(N_0 D_+-N_0 D_-)=\frac{(-e)}{4\pi}\tau\Delta,
\end{equation}
with $D_{\pm}=k_{F\pm}^2\tau/(2m^2)$. One sees that the longitudinal spin conductivity is simply  the difference between the Drude conductivities of the two spin channels and vanishes in the absence of uniform spin polarization at $\Delta\to 0$.

By replacing the spin current vertex $\hat J_x^x$ in Eq.~(\ref{14}) by $\hat J_y^y$, one can calculate the transverse spin conductivity $\sigma^{y}_{yx}$ from the same diagram as
\begin{equation}
\label{18}
\sigma^{y}_{yx} =\frac{1}{2\pi}\sum_{\kv} {\rm Tr} \left( J^y_y \hat G^R J_x\hat G^A\right),
\end{equation}
where the spin current vertex is given by
\begin{equation}
\label{19}
\hat J^y_y=\frac{1}{2}\frac{k_y}{m}\hat\sigma^y.
\end{equation}
Unfortunately, we find that $\sigma^{y}_{yx}$ from Eq.~(\ref{18}) vanishes after the trace over the Pauli matrices. This forces us to go beyond the zero-th order approximation and consider the velocity correction arising from the spin-orbit coupling with  impurities [diagrams in Fig.~\ref{diagram_2}(b) and (c)] as well as vertex corrections [diagrams in Fig.~\ref{diagram_2}(d) and (e)]. Explicit expressions for these diagrams are given  in Appendix A.  Specifically, the last term of our Hamiltonian in Eq.~(\ref{Hamiltonian}) gives rise to an anomalous velocity operator 
\begin{equation}
\label{20}
\delta \hat v_x \equiv \delta \hat v_{x,\kv,\kv'}={\rm i} \lambda^2 (k_{y}-k_{y}')\hat\sigma^z v_0.
\end{equation}
Note that the impurity-induced correction at the spin current vertex $\hat J_y^y$ is irrelevant because of the vanishing anti-commutator between $\hat\sigma^y$ and $\hat\sigma^z$.
The presence of $\hat\sigma^z$ together with the matrix structure
of the Green function allows to get an effective vertex which behaves as $\hat\sigma^y$ and then survives
when traced with $\hat J^y_y$. The diagrams in Figs.~\ref{diagram_2}(b) and (c), evaluated by standard techniques, yield
\begin{equation}
\label{21}
\sigma^y_{yx}(b+c)=e \frac{N_0 D}{2}\frac{2m\lambda^2 \Delta}{1+\Delta^2 \tau^2}\,.
\end{equation}
Further contributions arising from vertex corrections are shown in  Figs.~\ref{diagram_2}(d) and (e), where the impurity line connects a simple impurity potential insertion (full dot) with the  spin-orbit field due to the impurity (cross). 
The right part of those diagrams (including the impurity line) can be seen as a correction of the charge current vertex
\begin{equation}
\label{22}
\delta \hat J_x^{VC}=-2 m\lambda^2 \Delta \hat J^y_y,
\end{equation}
where the superscript ``VC'' stands for vertex corrections.  Evaluating the diagrams of Figs.~\ref{diagram_2} (d) and (e) according to the formulas given in Appendix A yields
\begin{equation}
\label{23}
\sigma^y_{yx}(d+e)=e\frac{N_0 D}{2}\frac{2m\lambda^2 \Delta}{1+\Delta^2\tau^2} ,
\end{equation}
which exactly matches the contribution from Figs.~\ref{diagram_2}(b) and (c) given by Eq.~(\ref{21}). 
The complete result can be cast in the form
\be
\label{HSHE1}
\sigma^y_{yx}=en\lambda^2\frac{\Delta \tau}{1+\Delta^2\tau^2}\,,
\ee
where we have made use of the relation $n=k_F^2/(2\pi)$ between density and Fermi wave vector to zero-th order in the spin-orbit coupling. 
This result has a simple and appealing physical interpretation: the prefactor $e n \lambda^2$ is simply the 
side-jump spin Hall conductivity, connecting the spin current $J^z_y$ to the electric field $E_x$.  The other factor $\Delta \tau/(1+\Delta^2\tau^2)$ gives the angle of rotation of the spin current about the direction of the spin polarization.  The expression for this angle agrees with the well known expression for the rotation of the equilibrium magnetization in the Hanle effect~\cite{Hanle1924} (see also Fig.~\ref{Hanle-SHE}).  Hence, we conclude that Eq.~(\ref{HSHE1}) is the  mathematical expression of the Hanle spin Hall effect when the spin Hall conductivity is evaluated in the Born approximation, which yields the so-called side-jump conductivity. Notice that, at variance with the Hanle effect for spin polarization~\cite{Dyakonovbook08,Ganichev02,Averkiev06},  it is the relaxation time of the spin current, approximately given by $\tau$, that enters Eq.~(\ref{HSHE1})  in lieu of the spin relaxation time.

Clearly, the spin current calculated from these diagrams should be observable in an experiment performed in the simple homogeneous setup described above, in which the electric current is a pure drift current.  Interestingly, the $J^y_y$ current generated through the HSHE is formally indistinguishable from the spin current generated by SCS, even though the physical origins of the two effects are completely different:  the HSHE depends crucially on the presence of the magnetic field to rotate the orientation of the spin current, while the   SCS does not.

\begin{figure}
\begin{center}
\includegraphics[width=2in]{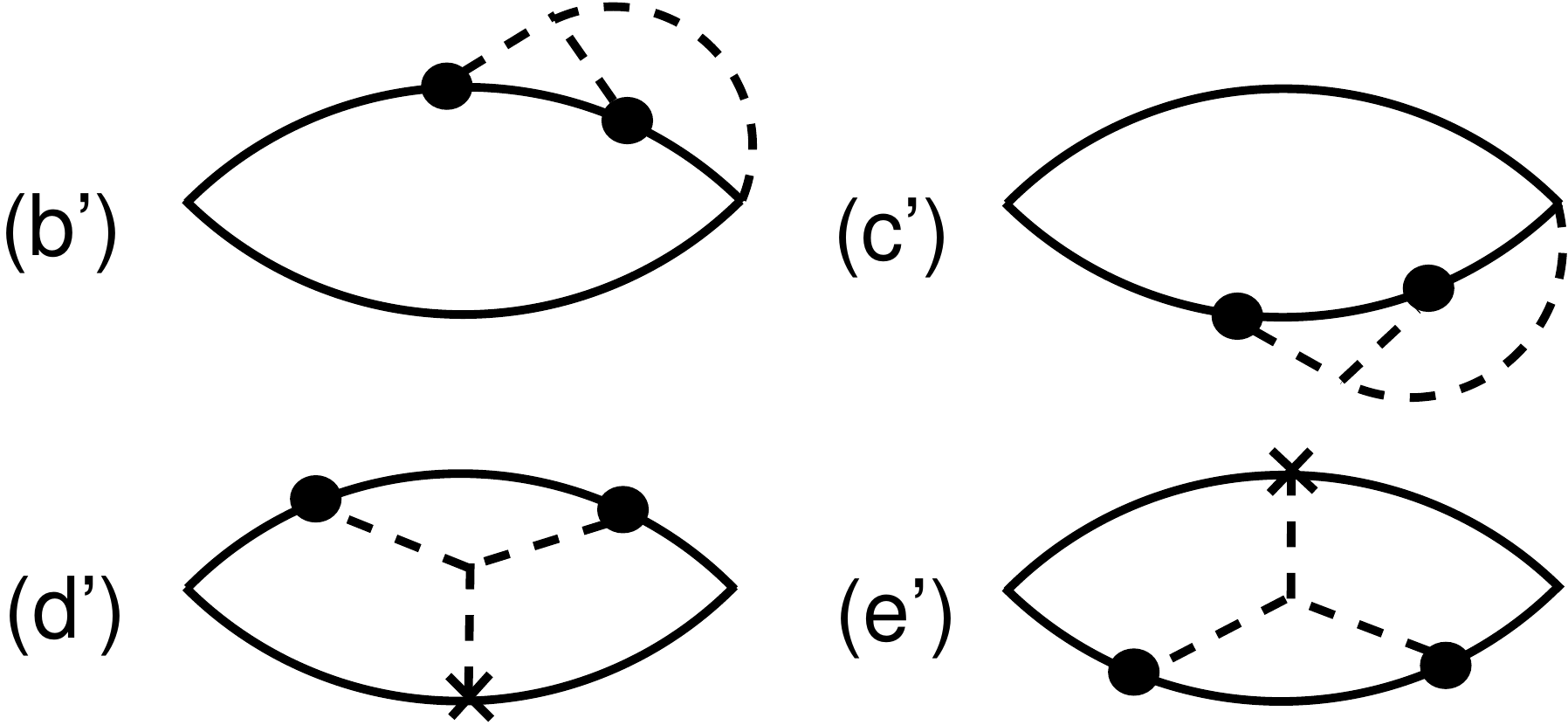}
\caption{Higher order corrections to the diagrams for the evaluation of the swapping coefficient.}
\label{diagram_2b}
\end{center}
\end{figure}

Up to this point we have limited ourselves to the lowest non-vanishing order (the second) in the impurity potential: this is why our Eq.~(\ref{HSHE1}) for the HSHE captures only the side-jump part of the spin Hall spin current.  However, the structure of this formula does not change when higher order diagrams are taken into account.  For example, in Fig.~\ref{diagram_2b}  we have considered the skew-scattering-like diagrams involving three impurity lines.   In analogy with the diagrams (b)+(c) and (d)+(e) of Fig.~\ref{diagram_2} we now have the diagrams  (b')+(c') and (d')+(e').  The new diagrams have the same structure as the ``parent diagrams" of  Fig.~\ref{diagram_2} and can all be obtained from the former  through the replacement 
\be
\label{amp_ren}
v_0 \to v_0 \sum_{\pv} G^{R(A)}_{\pv} v_0 \equiv \delta v^{R(A)}\,,
\ee
which is nothing but the first correction beyond the Born approximation to the spin-independent part of the scattering amplitude.  Hence one can combine second order and third order diagrams by introducing the renormalized scattering amplitudes
\begin{equation}
\label{AppB02}
v^{R(A)}=v_0+\delta v^{R(A)}=v_0\mp i \pi N_0 v^2_0.
\end{equation}
As discussed in Appendix \ref{APPB}, the contributions of the diagrams with renormalized amplitudes can be divided into two parts proportional to the two combinations $v^R+v^A$ and $v^R-v^A$, respectively. The former would give, in principle, no more than a renormalization of the scattering time in Eq.~(\ref{HSHE1}).  However, up to the third order we are considering, the scattering time is not renormalized, since the second order corrections in the scattering amplitudes cancel in the combination $v^R+v^A$, as it is clear from Eq.(\ref{AppB02}).
The latter gives rise to an additional contribution shown in Eq.~(\ref{AppB05}). To elucidate the meaning of this additional contribution, we recall that 
 the scattering amplitude in the presence of spin-orbit coupling can be written as\cite{Lifshits09}
\begin{equation}
\label{AppB2}
S=A+ B~{\hat {\bf k}}\times {\hat {\bf k}'}\cdot {\boldsymbol\sigma}\,,
\end{equation}
where, up to second order in perturbation theory in $v_0$,
but to first order in $\lambda^2$,  one has
\footnote{Notice that the evaluation to all orders in the scattering potential, but still to lowest order in the spin-orbit coupling does not change qualitatively this result.\cite{wolfle2006}}
\begin{equation}
\label{AppB4}
A=v_0-i\pi N_0 v_0^2, \   \   \  B=-i\lambda^2 k_F^2 v_0.
\end{equation}
According to  Ref.~\onlinecite{Lifshits09}, the combination  $   v^R-v^A \sim {\rm Re}(AB^*) $  corresponds to skew scattering processes. We therefore interpret Eq.~(\ref{AppB05}) as the skew-scattering part of the Hanle spin Hall effect.  And indeed, simple manipulations, shown in Appendix~\ref{APPB},  lead to the conclusion that the new term can be written as
the skew-scattering spin Hall conductivity times the ``Hanle factor" $\Delta \tau/(1+\Delta^2\tau^2)$. The final result, combining both side-jump and skew scattering contributions to the spin Hall conductivity  is
\be
\label{HSHE-final}
\sigma^y_{yx}=\sigma^z_{yx}\frac{\Delta \tau}{1+\Delta^2\tau^2}\,,
\ee
where
\be
\sigma^z_{yx}=ne\lambda^2\left(1+\frac{k_F^2v_0\tau}{4}\right)
\ee

 The conclusion is that the HSHE should be observable in the homogeneous experimental set up described above, with only a drift current and uniform in-plane electric and magnetic fields.    In contrast to this, an experiment that is optimally designed to observe the SCS should avoid both electric and magnetic fields in the conducting channel in which the effect is to observed.  We now turn to this question.

\section{Drift-diffusion equations and the SCS}  In order to show the role of the electric-field-induced spin-orbit coupling in a more apparent way, we now turn to the spin-dependent drift-diffusion equations. Beyond the simple model used above, in this section, we extend our discussion into more general cases by taking into account (i) the inhomogeneity of electronic spin density and (ii) spin-orbit coupling of ``intrinsic" origin, i.e., not related to the impurities.   The Hamiltonian can now be written in the SU(2) form
\be
H=\frac{\hat p^2}{2m}+\frac{1}{m}\hat p_i A_i^j\hat\sigma^j+e\mathbf E\cdot \mathbf r+V(\mathbf r)-\lambda^2 \vec{\hat\sigma}\times\nabla V(\mathbf r)\cdot \hat {\mathbf p},
\ee
where the SU(2) gauge field $A_i^j$ includes not only the intrinsic spin-orbit coupling, but also the one due to external electric field.  For example, in a $(001)$ two dimensional quantum well we have  $A_x^y=m(\alpha+\beta)$ and $A_y^x=m(\beta-\alpha)$ with $\alpha$ and $\beta$ corresponding to the coefficients of Rashba and Dresselhaus spin-orbit couplings separately.  In addition, the  in-plane electric field gives $A_x^z=\lambda^2 emE_y$ and $A_y^z=-\lambda^2emE_x$.

The conventional SU(2) drift-diffusion equation for the  spin current (defined as $J_i^a=\{\hat v_i,\hat\sigma^a\}/4$) reads~\cite{GoriniPRB10,Raimondi_AnnPhys12}
\be
J_{i}^{a}=-[(v_{i}+D\nabla_{i})S]^{a}-\theta_{\rm SH}\epsilon^{ija} J_j,
\label{spincurrent_dd}
\ee
where the last term on the right-hand side describes the spin-Hall term with $\epsilon^{ija}$ being the Levi-Civita antisymmetric tensor. Here, $v_i=e\tau E_i/m$ represents the drift velocity due to the external electric field. The covariant derivative $(\nabla_i O)^a=\partial_i O^a-2\epsilon^{abc} A_i^b O^c$.  However, as we noticed in Ref.~\onlinecite{Shenprb14},   the spin precession due to spin-orbit coupling with impurities is not  included in Eq.~(\ref{spincurrent_dd}). This effect can be derived  from the collision integral~\cite{Shenprb14}
\begin{equation}
I_{\mathbf k}(t)=-\left(\int_c dt^\prime[\Sigma_{\mathbf{k}}(t,t^\prime)G_{\mathbf{k}}(t^\prime,t)-G_{\mathbf{k}}(t,t^\prime)\Sigma_{\mathbf{k}}(t^\prime,t)]\right)^{<},\label{collisionG}
\end{equation}
with  the second-order self-energy
\be
\Sigma_{\kv}
= -in_{i}v_0^2\lambda^2\sum_{\mathbf{k}'}[{\vec{\hat\sigma}}\cdot\mathbf{k}\times\mathbf{k}',G_{\mathbf k'}].
\ee
Here, $G_{\kv}(\rv,t,t')$ and $\Sigma_{\kv}(\rv,t,t')$ stand for the contour-ordered Green's function and the self-energy, respectively. The superscript ``$<$'' denotes the lesser component of the contour integral. Since the detailed technique to calculate Eq.~(\ref{collisionG}) has been presented in Ref.~\onlinecite{Shenprb14}, here we jump to the result
\be
\label{Iscs}
I_{\mathbf{k}}^{\rm SCS} = -i\lambda^2({2\pi}{\tau})^{-1}\int_0^{2\pi} {d\theta_{\mathbf k'}}[{\vec{\hat\sigma}}\cdot\mathbf{k}\times\mathbf{k}^{\prime},\rho_{\mathbf{k}^{\prime}}]
\ee
where $\theta_{\mathbf k'}$ is the angle between $\mathbf k$ and $\mathbf k'$, and $\rho_{\mathbf k'}=\sum_i g^i_{\mathbf k'}\hat\sigma^i$ is the spin-dependent density matrix at momentum $\mathbf k'$.
In the steady state, Eq.~(\ref{Iscs}) leads to the following correction to the spin-dependent density matrix:
\be
\delta g_{\bf k}^{j}=(2\lambda^2m/N_0)\sum_{lmn}\epsilon^{zlj}\epsilon^{zmn}k_m J_n^l,
\ee
where the spin currents on the right-hand side are the ``unperturbed" ones:  $J_n^l\simeq\sum_{\mathbf k'} (k^\prime_n/m) g_{\bf k'}^l$. Then the additional contribution in spin current due to $\delta g_{\bf k}^{j}$ can be evaluated via
\be
[J_i^j]^{\rm SCS}\simeq \sum_{\mathbf k} (k_i/m) \delta g_{\bf k}^{j}=\kappa\left(J_j^i-\delta_{ij} J_l^l\right),
\ee
whose symmetry is consistent with previous work.\cite{Lifshits09} Here, the coefficient of SCS  reads $\kappa=\lambda^2 k_F^2$, as anticipated in the introduction. By adding this contribution to Eq.~(\ref{spincurrent_dd}), the complete spin current is  expressed as
\be \label{complete_spincurrent_dd}
J_{i}^{a}=-[(v_{i}+D\nabla_{i})S]^{a}+\kappa(J_{a}^{i}-\delta_{ia}J_{l}^{l})-\theta_{\rm SH}\epsilon^{ija} J_j\,.
\ee
One notices that $J_x^x$ and $J_x^y$ are coupled with $J_y^y$ and $J_y^x$ separately, while the spin Hall term does not contribute to the expressions for these components of the spin current:
\ber
J_{x}^{x}&=&-(v_{x}+D\partial_{x})S^{x}+2D\epsilon^{xbc}A_{x}^{b}S^{c}-\kappa J_{y}^{y},\label{Jxx1}\\
J_{y}^{y}&=&-(v_{y}+D\partial_{y})S^{y}+2D\epsilon^{ybc}A_{y}^{b}S^{c}-\kappa J_{x}^{x}.\label{Jyy1}
\eer
The first two terms on the right-hand side in each equation can be recognized as primary spin currents. Naturally, we can define the drift part of the spin currents as a product of the drift velocity and spin density, i.e.,
\ber
(J_{x}^{x})^{{\rm drift}}&=&	-v_{x}S^{x},\\
(J_{y}^{y})^{{\rm drift}}&=&	-v_{y}S^{y}.
\eer
The other part resulting from the diffusion effect can be written as
\ber
(J_{x}^{x})^{{\rm diff}}&=&	-D\partial_{x}S^{x}+2Dm(\alpha+\beta)S^{z},\label{Jxxdiff}\\
(J_{y}^{y})^{{\rm diff}}&=&	-D\partial_{y}S^{y}+2Dm(\alpha-\beta)S^{z}.\label{Jyydiff}
 \eer
One can see that in addition to the spatial inhomogeneity of the in-plane spin polarization, the out-of-plane spin polarization also contributes to the spin currents.   This contribution comes from the spin precession under the intrinsic spin-orbit effective magnetic field. Then, Eqs.~(\ref{Jxx1}) and (\ref{Jyy1}) can be rewritten as
\ber
J_{x}^{x}&=&(J_{x}^{x})^{{\rm drift}}+(J_{x}^{x})^{{\rm diff}}+\kappa(J_{y}^{y})^{{\rm drift}}-\kappa J_{y}^{y},\label{Jxx2}\\
J_{y}^{y}&=&(J_{y}^{y})^{{\rm drift}}+(J_{y}^{y})^{{\rm diff}}+\kappa(J_{x}^{x})^{{\rm drift}}-\kappa J_{x}^{x}.\label{Jyy2}
\eer
The third term on the right-hand side in each equation is obtained by substituting the vector potential $A_{x,y}^z$ into Eqs.(\ref{Jxx1}) and (\ref{Jyy1}), i.e., by taking into account the spin-orbit coupling due to the electric field.  The appearance of this term reduces the efficiency of the SCS effect.  Indeed, the equations show clearly that only the diffusion part of the primary spin current is a source of SCS.
By solving these equations, we obtain
\ber
J_{x}^{x}&=&(J_{x}^{x})^{{\rm drift}}+\frac{1}{(1-\kappa^2)}[(J_{x}^{x})^{{\rm diff}}-\kappa (J_{y}^{y})^{{\rm diff}}],\label{eq39}\\
J_{y}^{y}&=&(J_{y}^{y})^{{\rm drift}}+\frac{1}{(1-\kappa^2)}[(J_{y}^{y})^{{\rm diff}}-\kappa (J_{x}^{x})^{{\rm diff}}],\label{eq40}
\eer
from which we see that (i) the  drift component of the primary spin current $J_x^x$  does not generate SCS; (ii) a transverse spin current, $J_y^y$, is generated from the diffusive component of $J_x^x$ via SCS. The final expressions for the remaining spin currents, $J_y^x$ and $J_x^y$, can be obtained by simply replacing $J_x^x$, $J_y^y$ and $\kappa$ by $J_y^x$, $J_x^y$ and $(-\kappa)$, respectively.
\begin{figure}
\includegraphics[width=3in]{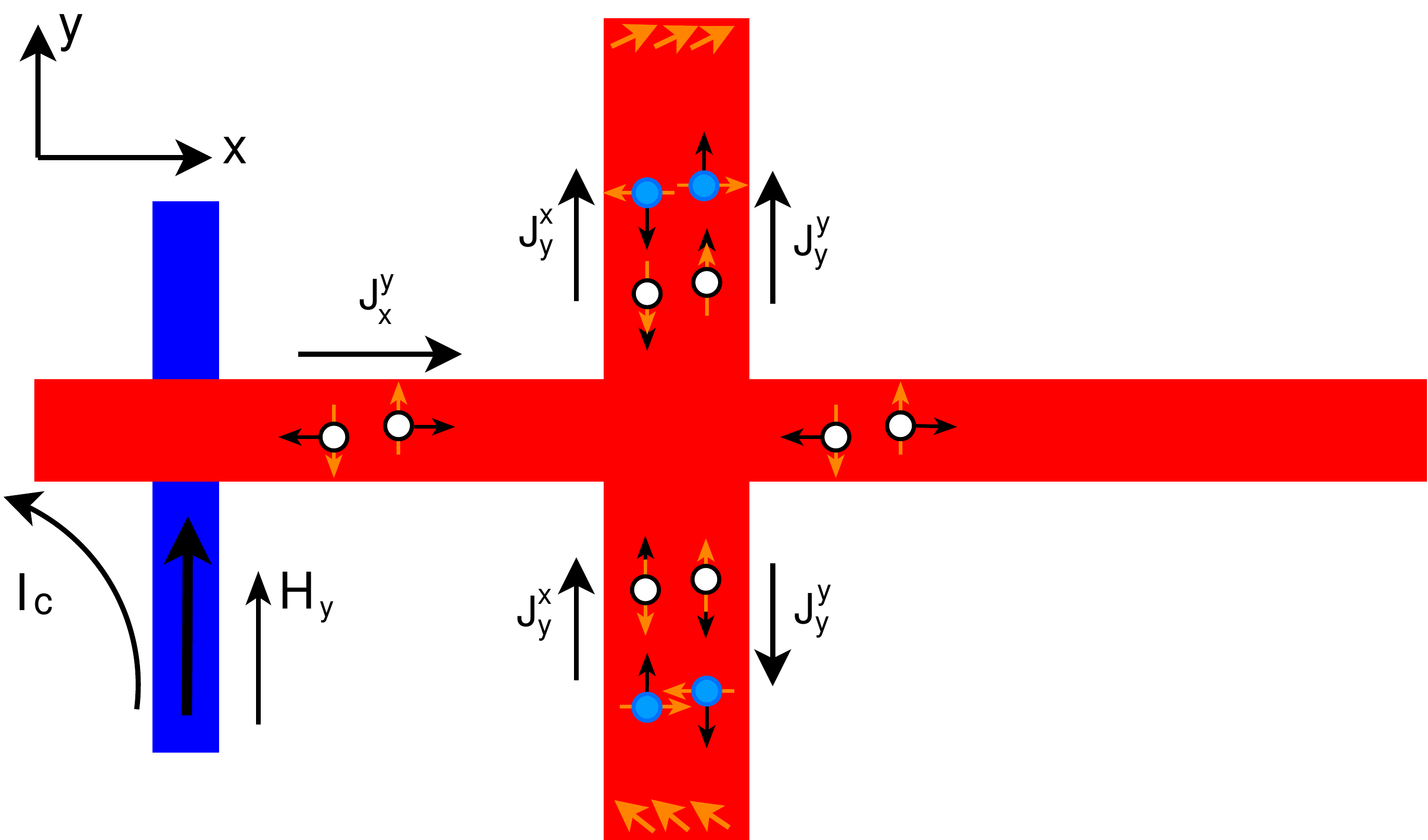}
\caption{Proposal on experimental observation of the spin current swapping effect. The spin polarization $S^y$ is accumulated at the cross area between ferromagnetic electrode (blue) and non-magnetic system (red) by electric current $I_c$ on the left circuit. Spin diffusion effect creates a spin current $J_x^y$  in the horizontal arm and $J^y_y$ in the transverse arms of the cross. In the addition,  spin current swapping  produces a spin current  $J_y^x$  in the transverse arms, resulting in different spin accumulations $S^x$ at the two ends  of the latter (shown as orange arrows).  The white  bubbles in the propagating channels illustrate the flow of the primary spin current density,  while the blue bubbles illustrate the flow of the secondary spin current arising from spin current swapping. In both cases the black arrows represent the direction of the current flow and the orange arrows the direction of the spin polarization.  
}
\label{exp_sugg}
\end{figure}

The SCS effect should be observable in an experiment such as the one described in Fig.~\ref{exp_sugg}, {in which intrinsic spin-orbit coupling is absent}.  The idea is to inject a pure spin current $J^y_x$ from a ferromagnetic contact into the longitudinal  ($x$) arm of a cross-shaped device.  Spin current swapping then injects a spin current $J^x_y$  into the transverse ($y$) arm of the cross resulting in opposite spin accumulations at the ends of the transverse arm.   These spin accumulations could in principle be detected by Faraday rotation spectroscopy (if the cross is made of a semiconductor material) or by inverse spin Hall effect (for metals).   In addition to the spin swapping current there is also a spin current $J^y_y$ flowing along the cross arm, originating from the diffusion of $y$-oriented spins from the center of the cross into the transverse arm.  This diffusion current produces equal spin accumulations on the two ends of the transverse arm and therefore does not contribute to the asymmetry.  We also notice that the inverse spin Hall effect associated with the primary spin current does {\it not} generate a potential difference between the ends of the transverse arm.\\

\section{Extrinsic versus intrinsic SCS}
{
In the presence of intrinsic spin orbit coupling, the spin accumulation at the edges of the transport channel in Fig.~\ref{exp_sugg} may become much more complicated and asymmetric features can show up even at $\kappa=0$. For example, according to Eq.~(\ref{complete_spincurrent_dd}), the injected spin $S^y$ can produce a spin current $J_y^z\simeq 2Dm(\beta -\alpha) S^y$, which makes the out-of-plane spin component $S^z$ accumulate at the two lateral edges with opposite sign, hence generates diffusion spin currents $[J_x^x]^{\rm diff}$ and $[J_y^y]^{\rm diff}$ according to Eqs.~(\ref{Jxxdiff}) and (\ref{Jyydiff}) and modifies the final spin accumulation map. Such phenomenon, induced solely by the intrinsic spin precession, was studied and named ``intrinsic spin current swapping'' by Sadjina et al.~\cite{Sadjina12}. 

In fact, these ``intrinsic" effects are implicit in the drift-diffusion equations as reported, for example, in Ref.~\onlinecite{Shenprb14},  but they do not show up as an explicit swapping term in those equations: this is why we could say that the spin-current swapping term {\it in the drift diffusion equations} has a
purely extrinsic origin --  its explicit form being given by the second term on the right hand side of Eq.~(\ref{complete_spincurrent_dd}).
Thus, if we look at the final outcome of any complete calculation of the spin current, we
expect to find both intrinsic and extrinsic contributions to what we call the spin-current-swapping spin current.  
But if we look at the equations themselves, there is only one explicit spin current swapping
term, and that is the extrinsic one -- the second term on the right hand side of Eq.~(\ref{complete_spincurrent_dd}).

Obviously, for a comprehensive calculation of the local spin accumulation in the transport channel, one needs to self-consistently solve the complete spin drift-diffusion equations (see Ref.~\onlinecite{Shenprb14}) including both spin precession and spin current swapping with proper boundary condition as was done in Ref.~\onlinecite{Sadjina12},  where the $\kappa$ term is, however,  missing.  From an experimental point of view, it is a big challenge to distinguish the contribution of the extrinsic spin current swapping ($\kappa$ term) from the intrinsic spin precession.  This is why  to observe the direct influence of the spin current swapping term on the spin accumulation it is better to perform the measurement in a system in which the intrinsic spin precession is negligible, as suggested in the previous section.
 }

\section{Acknowledgements} Ka Shen and Giovanni Vignale acknowledge support  from NSF Grant No. DMR-1406568.\\

\appendix
  \section{Hanle spin Hall effect -- the side-jump contribution}
  Up to the first order in $\lambda^2$, the transverse spin conductivity $\sigma_{yx}^{y}$ from side-jump-like diagrams, corresponding to Figs.~\ref{diagram_2}(b) and (c), is given by
\ber
\sigma_{yx}^{y}(b+c)&=&ie\lambda^2 n_{i}v_{0}^{2}\frac{1}{2\pi}\sum_{\mathbf{k}\mathbf{k}'}\frac{k_{y}}{2m}(k_{y}-k_{y}')\nonumber\\
      &&\hspace{0.5cm}\times{\rm Tr}[\sigma^{y}G_{\mathbf{k}}^{R}(G_{\mathbf{k}'}^{R}\sigma_{z}-\sigma_{z}G_{\mathbf{k}'}^{A})G_{\mathbf{k}}^{A}].
      \label{lowest_sj}
\eer
The impurity-vertex-correction diagrams, i.e., Figs.~\ref{diagram_2}(d) and (e), lead to
\ber
\sigma_{yx}^{y}(d+e)&=&ie\lambda^2 n_{i}v_{0}^{2}\frac{1}{2\pi}\sum_{\mathbf{k}\mathbf{k}'}\frac{1}{2}\frac{k_{y}}{m}\frac{k_{x}'}{m}(k_{x}k_{y}'-k_{y}k_{x}')\nonumber\\
      &&\hspace{0.5cm}\times{\rm Tr}[\sigma^{y}G_{\mathbf{k}}^{R}[\sigma_{z},G_{\mathbf{k}'}^{R}G_{\mathbf{k}'}^{A}]G_{\mathbf{k}}^{A}].
\eer

 \section{Hanle spin Hall effect -- the skew-scattering contribution}
\label{APPB}
The contribution of diagrams (b') and (c') of Fig. \ref{diagram_2b} reads
\begin{eqnarray}
\nonumber
\sigma^y_{yx}(b'+c')&=&ie\lambda^2n_i v_0\frac{1}{2\pi}\sum_{{\bf k}{\bf k'}}\frac{k_y^2}{2m}{\rm Tr}\left[G^A_{\bf k}\hat \sigma^y G^R_{\bf k} \right.\\
&&\left. 
(\delta v^R G^R_{{\bf k'}} \hat\sigma^z -\hat\sigma^z G^A_{{\bf k'}}\delta v^A )
\right]\,,\label{AppB01}
\end{eqnarray}
where $\delta v^{R}$ and $\delta v^{A}$ are defined in Eq.~(\ref{AppB02}). By performing the sum over ${\bf k'}$ only the imaginary part remains with an  opposite sign for retarded and advanced Green's functions.
Because of the anticommutation property of the Pauli matrices $\sum_{\bf k'}(\delta v^R G^R_{{\bf k'}} \hat\sigma^z -\hat\sigma^z G^A_{{\bf k'}} \delta v^A)=-i\pi N_0 (\delta v^R+\delta v^A)\sigma_z\simeq0$ and Eq.~(\ref{AppB01}) vanishes.

Diagrams (d') and (e') can be analyzed similarly. Their expression reads
\begin{eqnarray}
\nonumber
\sigma^y_{yx}(d'+e')&=&-i e\lambda^2 n_i v_0 \frac{1}{2\pi}\sum_{{\bf k}{\bf k'}} \frac{k_y ^2 k_x'^2}{2m^2}{\rm Tr} \left[ G^A_{\bf k}\hat \sigma^y G^R_{\bf k} \right.\\
&&\left. 
(\hat\sigma^z G^R_{\bf k'}G^A_{\bf k'} \delta v^A-\delta v^RG^R_{\bf k'}G^A_{\bf k'}\hat\sigma^z)
\right].\label{AppB03}
\end{eqnarray}
By using the following  identity valid for any two operators $M$ and $N$
\begin{equation}
\label{AppB04}
\delta v^A M N-\delta v^R N M=
\frac{\delta v^R+\delta v^A}{2} \left[ M,N \right]+ \frac{\delta v^A-\delta v^R}{2} \big\{ M,N\big\},
\end{equation}
one may see that the contribution (\ref{AppB03}) splits in two terms proportional to the combinations $\delta v^R+\delta v^A$ and $\delta v^R-\delta v^A$. The former vanishes due to the fact $\delta v^R+\delta v^A\sim 0$. The latter contribution of Eq.(\ref{AppB03}), proportional to $\delta v^R-\delta v^A$, remains and
yields
\ber
\label{AppB05}
\sigma^y_{yx, \rm res}&=&e\lambda^2 \frac{n_i}{2\pi}(mv_0)^3\frac{\tau^3\Delta E_F^2}{1+\Delta^2\tau^2}.\nonumber\\
&=&en\lambda^2 \frac{v_0k_F^2\tau}{4} \frac{\tau\Delta}{1+\Delta^2\tau^2}
\eer
where the prefactor $en\lambda^2 v_0k_F^2\tau/4$ equals the spin Hall conductivity due to skew scattering.\cite{Tse06}
As explained in the main text this contribution proportional to $\delta v^R-\delta v^A$ reflects the origin of the skew-scattering processes\cite{Lifshits09}, which
give rise to the spin Hall effect and, in the presence of an exchange field,
leads to a coupling between the two spin currents $J^y_y$ and $J^z_y$ and hence to 
the residual $\sigma^y_{yx,\rm res}$ of Eq.(\ref{AppB05}). 

To illustrate how the combined action of the skew scattering and the exchange field leads to the above additional contribution to the $J^y_y$ spin current, let us consider the skew scattering  contribution to the collision integral $I^{\rm ss}_{\bf k}$,
which was derived in Eq.(24) of Ref.\onlinecite{Shenprb14}
\ber
\label{AppB1}
I^{\rm ss}_{\bf k}&=&n_i \lambda^2\frac{m^2v_0^3}{2}\langle\{ {\bf k}\times {\bf k'}\cdot {\boldsymbol \sigma}, \rho_{\bf k'}\}\rangle.\nonumber \\
&=&2\pi N_0 n_i{\rm Re}(AB^*)\langle \{ {\hat {\bf k}}\times {\hat {\bf k}'}\cdot {\boldsymbol \sigma}, \rho_{\bf k'}\} \rangle
\eer
where $\langle \dots \rangle \equiv (2\pi)^{-1}\int d\theta_{\bf k'}$. By generalizing the kinetic equation developed in Ref. \onlinecite{Shenprb14} (cf. its Eq.(13)) in the presence of magnetic field, we obtain (keeping only the skew scattering besides the standard scattering)
\begin{equation}
\label{AppB6}
-i\frac{\Delta}{2}\left[\hat \sigma^x,\rho_{\bf k}\right]-e{\bf E}\cdot \nabla_{\bf k}\rho_{\bf k}=-\frac{1}{\tau}
(\rho_{\bf k}-\langle \rho_{\bf k'}\rangle)+I^{\rm ss}_{\bf k}.
\end{equation}
After projecting the kinetic equation along the $\hat \sigma^z$ and $\hat\sigma^y$ components and considering the ${\hat k}_y$ partial p-wave, one obtains the two coupled equations
\begin{eqnarray}
\Delta \langle {\hat k}_y g_{\bf k}^y\rangle&=&-\frac{1}{\tau}\langle {\hat k}_y g_{\bf k}^z\rangle
-n_i\lambda^2k_F^2 \frac{m^2v_0^3}{2} \langle {\hat k}_x g^0_{\bf k}\rangle \label{AppB7}\\
-\Delta \langle {\hat k}_y g_{\bf k}^z\rangle&=&-\frac{1}{\tau}\langle {\hat k}_y g_{\bf k}^y\rangle.\label{AppB8}
\end{eqnarray}
In the absence of the magnetic field ($\Delta =0$), by considering that $J^z_y\sim \langle {\hat k}_y g_{\bf k}^z\rangle$ and $J_x\simeq ne^2\tau E_x/m\sim -2e\langle {\hat k}_x g^0_{\bf k}\rangle$ one has the spin Hall effect. By switching on the magnetic field the spin current $J^y_y\sim \langle {\hat k}_y g_{\bf k}^y\rangle$ couples with $J^z_y$ and its value as obtained from Eqs.(\ref{AppB7}-\ref{AppB8}) coincides with 
Eq.(\ref{AppB05}).

\bibliography{srv_scswapping.bib}
\bibliographystyle{prsty}
\end{document}